\documentclass[12pt]{iopart}

\usepackage{iopams}  
\usepackage{graphicx}
\usepackage{amssymb}

\expandafter\let\csname equation*\endcsname\relax
\expandafter\let\csname endequation*\endcsname\relax
\usepackage{amsmath}

\usepackage{stackrel}
\usepackage{xcolor}
\usepackage{tikz}
\usetikzlibrary{arrows}

\begin{document}

\title[]{Deformable active nematic particles and emerging edge currents in circular confinements}

\author{Veit Krause}
\address{Institut für Wissenschaftliches Rechnen, TU Dresden, 01062 Dresden, Germany}

\author{Axel Voigt}
\address{Institut für Wissenschaftliches Rechnen, TU Dresden, 01062 Dresden, Germany and Center for Systems Biology Dresden (CSBD), Pfotenhauerstr. 108, 01307 Dresden, Germany and Cluster of Excellence - Physics of Life, TU Dresden, 01062 Dresden, Germany}
\ead{axel.voigt@tu-dresden.de}
\vspace{10pt}
\begin{indented}
\item[]January 2021
\end{indented}

\begin{abstract}
We consider a microscopic field theoretical approach for interacting active nematic particles. With only steric interactions the self-propulsion strength in such systems can lead to different collective behaviour, e.g., synchronized self-spinning and collective translation. The different behaviour results from the delicate interplay between internal nematic structure, particle shape deformation and particle-particle interaction. For intermediate active strength an asymmetric shape emerges and leads to chirality and self-spinning crystals. For larger active strength the shape is symmetric and translational collective motion emerges. 
Within circular confinements, depending on the packing fraction, the self-spinning regime either stabilizes positional and orientational order or can lead to edge currents and global rotation which destroys the synchronized self-spinning crystalline structure.  
\end{abstract}

%
%
%
%
%

\section{Introduction}

Active matter systems take energy from their environment and drive themselves out of equilibrium. This can lead to novel collective phenomena and provides hope to uncover the physics of living systems and to find new strategies for designing smart devices and materials. We refer to \cite{Ramaswamy_ARCMP_2010,Marchettietal_RMP_2013,Catesetal_ARCMP_2015,Menzel_PR_2015,Bechingeretal_RMP_2016,Gompperetal_JPCM_2020} for various reviews. An important example of active matter is constituted by natural and artificial objects capable of self-propulsion. A fundamental challenge is to understand how such objects interact and lead to collective phenomena. Most of the microscopic modeling approaches in this field consider active particles which have a fixed symmetric shape, and movement is defined along a symmetry axis. This leads to motion along a straight line just perturbed by random, e.g., Brownian fluctuations. Both assumptions, on shape and symmetry, are restrictive, as shape deformations as well as deviations from symmetry destabilize any straight motion and make it chiral, which would result in circular motion. As most systems are imperfect this should be the general case. While attempts exist to generalize active particle models in this direction, see, e.g., \cite{Liebchenetal_PRL_2017,Levisetal_PRE_2019,Levisetal_PRR_2019,Kruketal_PRE_2020} for imposed alignment mechanisms, \cite{Denketal_PRL_2016,Baeretal_ARCM_2020} for anisotropic particle shapes and \cite{Ohtaetal_PRL_2009,Menzeletal_EPL_2012} for shape deformations, multiphase-field models, e.g., \cite{Mueller_PRL_2019,Wenzel_JCP_2019,Loewe_PRL_2020}, where each object is modeled by a phase field variable, naturally allow for shape deformability and also provide the possibility to incorporate asymmetry to enforce chirality. It has already been demonstrated that collisions of deformable objects can lead to alignment \cite{Grossmanetal_NJP_2008,Menzeletal_EPL_2012,Loeberetal_SR_2015,Marthetal_IF_2016}. As a result, these multiphase-field models do not require any explicit alignment interactions. A drawback of such models is the huge computational effort for large numbers of interacting objects. We here consider an intermediate modeling approach. The approach considers a particle density for all particles and combines it with internal nematic structure. We are only interested in relatively dense systems and study the influence of activity in unconfined and confined domains.

The paper is organized as follows: In Section \ref{s:modeling} we postulate a minimal model which is capable of shape deformations and broken symmetry with respect to the direction of motion. The model is termed nematic active phase field crystal model. Besides the motivation, the evolution equations are explained and the numerical approaches for solving and postprocessing are sketched. Section \ref{s:single} analyses the model for a single object and identifies three different regimes: resting, circular or spinning motion and translation. Section \ref{s:collective} considers the emerging collective behaviour in unconfined and confined geometries, and Section \ref{s:collective} discusses these results and relates the observed phenomena to that of other theoretical and experimental investigations.

\section{Modeling}
\label{s:modeling}

\subsection{Motivation}

Microscopic field theoretical approaches for active system can be considered as a compromise between the full details of multiphase-field models and active particle models. They first have been introduced in \cite{Menzeletal_PRL_2013} for active crystals and consider a local particle density variation field $\psi$, a local polar particle orientation field $\mathbf{P}$ and a self-propulsion strength $v_0$. The model combines a phase field crystal model for freezing \cite{Elderetal_PRL_2002,Elderetal_PRE_2004} with a Toner-Tu model for self-propelled particles  \cite{Toneretal_PRL_1995}. More recently this approach was also considered on surfaces \cite{Praetoriusetal_PRE_2018} and has been extended by an active torque and the interplay of self-propulsion and self-spinning of crystallites was investigated in \cite{Huangetal_PRL_2020}. A different path was followed in \cite{Alaimoetal_NJP_2016,Alaimoetal_PRE_2018} where the underlying phase field crystal energy was modified to consider independent active particles \cite{Chanetal_PRE_2009,Berryetal_PRL_2011,Robbinsetal_PRE_2012,Ophausetal_PRE_2018}. The approach allows to simulate a transition from a resting particle to a moving state by increasing the self-propulsion strength. Within this transition, the particle deforms and elongates perpendicular to the direction of motion. Other phenomena, considered for more particles, are cell–cell collisions, oscillatory motion in confined geometries, collective migration and cluster formation in homogeneous systems \cite{Alaimoetal_NJP_2016}, as well as the rich dynamics of heterogeneous systems of active and passive particles, ranging from highly dilute suspensions of passive particles in an active bath to interacting active particles in a dense background of passive particles \cite{Alaimoetal_PRE_2018}. A common characteristic of these models is the presence of an underlying interaction potential for the density variation field $\psi$ but no enforcement of aligning the polar particle orientation field $\mathbf{P}$. Alignment results solely from inelastic particle deformations through their interaction \cite{Grossmanetal_NJP_2008}. 

\begin{figure}[htb]
   \centering
   \setlength{\unitlength}{\textwidth}
   \includegraphics[width=0.5\textwidth]{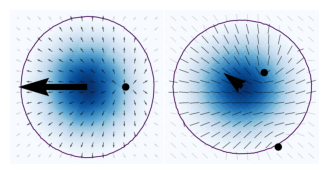}
   \caption{Particle density variation field $\psi$ for a single density peak (color coding) with the $0.01$ level-set indicating the shape of the particle. The internal polar (left) and nematic (right) structure is visualized by the director field. The figure further shows the direction and strength of motion (arrows) and the location of topological defects (black points). For the polar model (left) one $+1$ defect is located on the symmetry axis and for the nematic model (right) two $+1/2$ defects break the symmetry and make the motion chiral.}
   \label{fig1}
\end{figure}

For a single particle the interplay between a splay (or bent) instability of the polar particle orientation field $\mathbf{P}$, the particle shape and the strength of the self-propulsion $v_0$ has been discussed \cite{Alaimoetal_NJP_2016} and corresponds to the same mechanism as in phase-field models for active droplets, see, e.g. \cite{Tijungetal_PNAS_2012,Marthetal_JRSI_2014}. As a result of the vertical anchoring of the polar particle orientation field $\mathbf{P}$ at the particle boundary, one $+1$ defect forms within the particle. With no further interaction, due to the isotropic properties of the particle, the resulting shape of the particle is symmetric with respect to the direction of motion, see Fig. \ref{fig1}(a). To incorporate chirality thus requires an additional active forces, as in \cite{Huangetal_PRL_2020}, or a different particle orientation field. Adapting approaches of active nematic droplets \cite{Giomietal_PRL_2014,Gaoetal_PRL_2017}, we propose a microscopic field-theoretical approach, which couples a local particle density variation field $\psi$, a local nematic particle Q-tensor field $\mathbf{Q}$ and a self-propulsion strength $v_0$. Similar mechanisms, as described above, also follow for this model, but now the nematic properties lead to the presence of two $+1/2$ defects, which allows to break the symmetry and induces chirality, see Fig. \ref{fig1}(b). This property allows to consider only one active parameter, the self-propulsion strength $v_0$, to tune the rich dynamics of the model. 

\subsection{Evolution equations}

The proposed minimal model reads
\begin{align}
    \partial_t\psi &= M_0 \Delta \frac{\delta {\cal{F}}_{vPFC}}{\delta\psi} + v_0 \nabla \cdot (\psi \mathbf{Q}\nabla\psi)   \\
    \partial_t \mathbf{Q} &=  L \Delta \mathbf{Q} - c(\operatorname{tr} {\mathbf{Q}^2} - 1)\mathbf{Q} - v_0( 2\nabla\psi\nabla\psi^T - \Vert{\nabla\psi}\Vert^2Id ) -\beta 1_{\{ \psi>0 \}}\mathbf{Q}.
\end{align}
with particle density variation field $\psi$, nematic particle Q-tensor field $\mathbf{Q}$ and self-propulsion strength $v_0$. The first equation considers conserved dynamics for the free energy ${\cal{F}}_{vPFC} = {\cal{F}}_{PFC} + \int H (|\psi|^3 - \psi^3) \, d \mathbf{r}$, with
\begin{align}
    {\cal{F}}_{PFC} = \int \frac{\psi}{2} (r + (1 + \nabla^2)^2) \psi) + \frac{\psi^4}{4} \; d \mathbf{r}
\end{align}
the Swift-Hohenberg energy \cite{Swiftetal_PRA_1977,Elderetal_PRL_2002,Elderetal_PRE_2004}, with parameter $r$ related to an undercooling, and an additional penalization term, with parameter $H>0$. The penalization enforces the density variations to remain positive. This modifies the particle interaction and allows to phenomenologically describe independent particles \cite{Chanetal_PRE_2009,Berryetal_PRL_2011,Robbinsetal_PRE_2012,Ophausetal_PRE_2018}. 
A detailed derivation of ${\cal{F}}_{PFC}$ and its relation to classical density functional theory can be found in \cite{Elderetal_PRB_2007,Teeffelenetal_PRE_2009,Vrugtetal_AP_2020}. The variational derivative reads $\frac{\delta {\cal{F}}_{vPFC}}{\delta\psi} = (r + 1) \psi + 2 \nabla^2 \psi + (\nabla^2)^2 \psi + \psi^3 + 3H(\psi |\psi| - \psi^2)$. The parameter $M_0$ sets a mobility and is responsible for the deformability of the density peaks. The active contribution is considered in analogy to the polar model \cite{Alaimoetal_NJP_2016}, with $\mathbf{Q} \nabla \psi$ playing the role of the polar particle orientation field $\mathbf{P}$. The second equation considers unconserved dynamics of a Landau-de Gennes energy in its one-constant approximation
\begin{align}
    {\cal{F}}_{LdG} = \int \frac{L}{2} \| \nabla \mathbf{Q} \|^2 + \frac{a}{2} \operatorname{tr} \mathbf{Q}^2 + \frac{2}{3} b \operatorname{tr} \mathbf{Q}^3 + \frac{c}{4} \operatorname{tr} \mathbf{Q}^4 \; d \mathbf{r}
\end{align}
with elastic constant $L$ and entropic parameters $b = 0$ and $a = -c$. The active component is constructed to ensure the Q-tensor properties and the last term restricts, in analogy to \cite{Alaimoetal_NJP_2016}, the nematic particle Q-tensor field $\mathbf{Q}$ to be different from zero only within the particles, with $\beta > 0$.

\subsection{Numerical approach}

The coupled equations are reformulated as a set of second order equations and solved using an operator splitting approach for $\psi$ and $\mathbf{Q}$ in a semi-implicit manner. Discretisation in space is done by finite elements \cite{Backofenetal_PML_2007,Praetoriusetal_SIAMJSC_2015} and adaptive refinement is considered to ensure a fine discretisation within the particles. The approach is implemented in AMDiS \cite{Veyetal_CVS_2007,Witkowskietal_ACM_2015}.

We consider a square domain $\Omega = [-6d,6d]^2$, with periodic boundary conditions, where $d = \frac{4 \pi}{\sqrt{3}}$ the lattice distance of the phase field crystal model. A circular confinement is enforced using an interaction potential to be added to ${\cal{F}}_{vPFC}$, which reads $\int B \psi^2 \varphi_B \, d \mathbf{r}$ with $B>0$ and $\varphi_B$ a $\tanh$-approximation of $1_{\Omega \backslash \Omega_c}$, with $\Omega_c = \{ \| \mathbf{r} \| < 6d \}$. 

The model parameters are fixed as $r = -0.9$, $M_0 = 20$, $L = 0.2$, $c = 0.1$, $H = 10^5$, $\beta = 10$ and $B = 10^5$. The self-propulsion strength $v_0$ will be varied and specified below. Numerical parameters concerning grid resolution, time step and $\tanh$-approximation are chosen to guarantee mesh-independency and stable behaviour.  

As initial condition we specify
$ \psi_0 = A\sum_{i=1}^N (\cos(\frac{\sqrt{3}}{2}\Vert \mathbf{r} - \mathbf{r}_i \Vert)+1) 1_{\Vert \mathbf{r} - \mathbf{r}_i \Vert <2\pi/\sqrt{3} } $
with prefactor $A$ such that
$\int \psi_0 \; d \mathbf{r} = \frac{N d^2}{\vert\Omega\vert} \sqrt{(-48-56r) / 133}$ and particle initial positions $\mathbf{r}_i$ for $i = 1,\ldots,N$ with $N$ the number of particles. As initial Q-tensor field we consider a symmetric field with one $+1$ defect in the center of each particle and vertical anchoring at the particle boundary. The symmetric Q-tensor field is perturbed by white noise. The $+1$ defects are unstable and immediately split into two $+1/2$ defects. The way these defects rearrange sets the shape of the particle and its direction of movement.  

For postprocessing purposes the center of the $i$-th particle at time $t^n$ is computed as $\mathbf{r}_i^n = \int_{\mathcal{B}_i} \mathbf{r} \psi^n \; d \mathbf{r} /  \int_{\mathcal{B}_i} \psi^n \; d \mathbf{r}$, with $\mathcal{B}_i$ a small circle around the maximum of the $i$-th density peak. The radius of $\mathcal{B}_i$ is related to $d$. The $i$-th particle velocity follows as $\mathbf{v}_i^n = (\mathbf{r}_i^{n} - \mathbf{r}_i^{n-1})/(t^n- t^{n-1})$ and the mean particle velocity magnitude is the average over all $\mathbf{v}_i^n$, computed as $\overline{v}^n = \frac{1}{N} \sum_{i=1}^N \|\mathbf{v}_i^n\|$.   

As in \cite{Loeberetal_SR_2015,Alaimoetal_NJP_2016} we define for every time $t^n$ the translational order parameter $\phi_{T}^n$ and the rotational order parameter $\phi_{R}^n$ with
\begin{align*}
    \phi_{T}(t^n) = \frac{1}{N}\| \sum \limits_{i=1}^N \hat{\mathbf{v}}_{i}^n \|,
    \quad
    \phi_{R}(t^n) = \frac{1}{N} \sum \limits_{i=1}^N (\hat{\mathbf{r}}_{i}^n)^T\hat{\mathbf{v}}_{i}^n \quad \mbox{and}
    \quad
    \phi_{O}(t^n) = \sin (\frac{1}{N}{\sum \limits_{n=1}^N \arctan({\mathbf{v}_{i}^n)}} )
\end{align*}
where $\hat{\mathbf{v}}_{i}^n = \mathbf{v}_{i}^n / \|\mathbf{v}_{i}^n\|$ is the unit i-th particle velocity vector and $\hat{\mathbf{r}}_{i}^n = \mathbf{r}_{i}^n / \|\mathbf{r}_{i}^n\|$ the unit i-th particle position vector at time $t^n$. In case of collective translation or collective rotation, we get $\phi_{T,i} \approx 1$ or $|\phi_{R,i}| \approx 1$, respectively. However, also collective orientation in synchronously spinning particles leads to $\phi_{T,i} \approx 1$. To distinguish translational and orientational order $\phi_{O}$ measures synchronously changing orientation. The frequency of the oscillation in $\phi_O(t)$ determines the collective angular spinning velocity. 

\section{Single particle}
\label{s:single}

\begin{figure}[htb]
   \centering
   \setlength{\unitlength}{\textwidth}
   \includegraphics[width=0.8\textwidth]{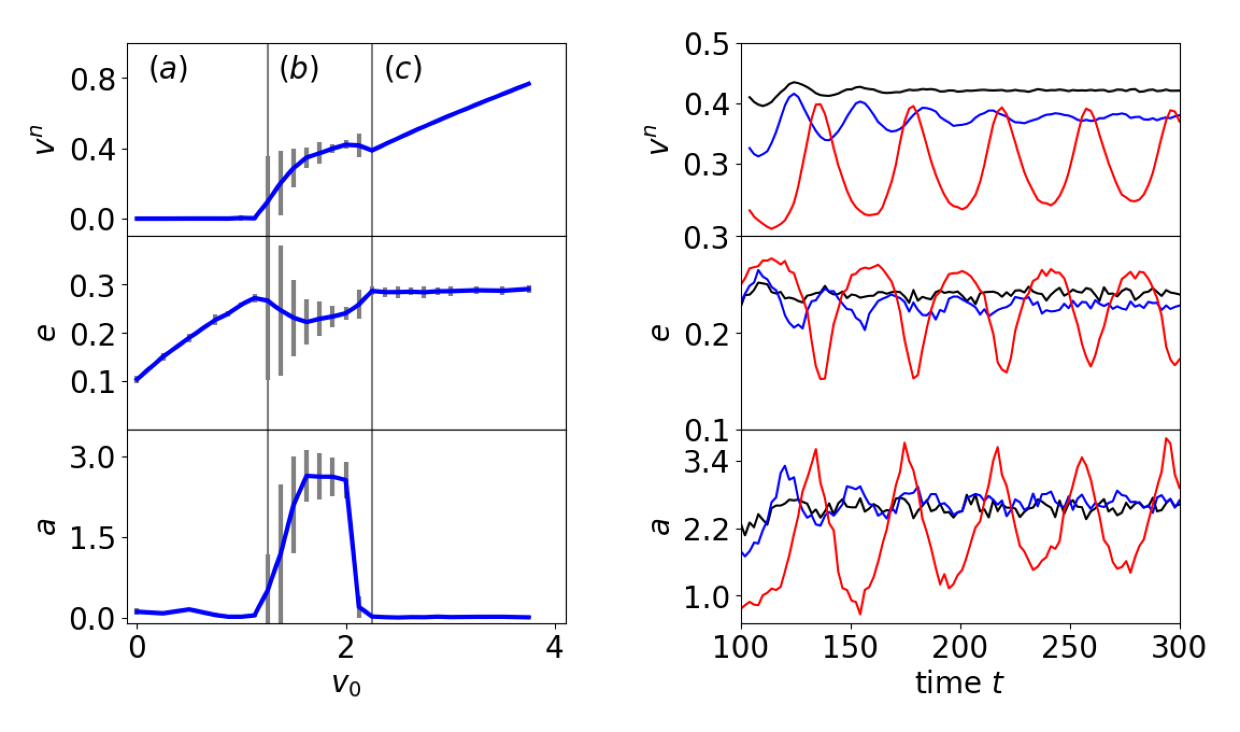}
   \caption{Particle velocity (top), eccentricity (middle) and defect asymmetry (bottom) of a single particle depending on self-propulsion strength $v_0$ (left). The vertical lines indicate the separation into three regimes. From left to right: (a) resting, (b) circular or spinning motion, and (c) translation. The error bars correspond to values at different times within the nonequilibrium steady state. For the circular or spinning regime the dynamically stable state is shown as a function of time (right) for $v_0 = 1.5$ (red), $v_0 = 1.75$ (blue) and $v_0 = 2.0$ (black).}
   \label{fig2}
\end{figure}

We first consider the situation of one particle. It is placed in the centre of the domain and we consider the effect of $v_0$. Fig. \ref{fig2} shows the particle velocity, the eccentricity and the asymmetry of the defect arrangement as a function of $v_0$. The eccentricity is defined as $e_i^n = \sqrt{1-(r_{i,min}^n)^2/(r_{i,max}^n)^2}$, where $r_{i,min}^n$ and $r_{i,max}^n$ are the minimal and maximal distances between the center of mass and the $0.15$-levelset of $\psi$ for particle $i$ at time $t^n$, respectively. The $0.01$-levelset is considered as the particle boundary. The asymmetry of the defect arrangement is computed as the deviation from the center of mass with respect to length and angle, as $a_i^n = \vert \Vert \mathbf{d}_{i,1}^n -\mathbf{r}_i^n\Vert - \Vert \mathbf{d}_{i,2}^n -\mathbf{r}_i^n\Vert\vert + \left\vert\frac{(\mathbf{d}_{i,1}^n-\mathbf{r}_i^n)^T\mathbf{v}_i^n}{\Vert \mathbf{d}_{i,1}^n-\mathbf{r}_i^n\Vert} - \frac{(\mathbf{d}_{i,2}^n-\mathbf{r}_i^n)^T\mathbf{v}_i^n}{\Vert \mathbf{d}_{i,2}^n-\mathbf{r}_i^n\Vert} \right\vert$, where $\mathbf{d}_{i,1}^n$ and $\mathbf{d}_{i,2}^n$ are the positions of the two $+1/2$ defects for particle $i$ at time $t^n$. Various approaches exist to determine defects in nematic liquid crystals, see \cite{wenzel2020defects} for a comparison of various methods. We here consider them as degenerate points of $\mathbf{Q}$ for which $Q_{11} = Q_{12} = 0$. This allows an easy detection of the position of a defect. For a nematic liquid crystal in 2D two types of topological defects predominate $+ 1/2$ and $- 1/2$. Considering the sign of $\delta = \frac{\partial Q_{11}}{\partial x} \frac{\partial Q_{12}}{\partial y} - \frac{\partial Q_{11}}{\partial y} \frac{\partial Q_{12}}{\partial x}$ allows to distinguish between them. Due to the setting within a particle and the specified vertical anchoring only $+1/2$ defects occur in the considered parameter regime. 

Fig. \ref{fig2}(left) shows three regimes: (a) resting, characterized by a zero velocity, the cell shape deforms with increasing $v_0$ and the defect positions are symmetric, (b) circular or spinning, the velocity fluctuates, which has an effect on the eccentricity and the asymmetry of the defect positions, and (c) translation, with increasing velocity, constant shape and symmetric arrangement of defects. The nonequilibrium steady state of the circular or spinning regime is shown in Fig. \ref{fig2}(right) for different $v_0$. The oscillations underpin the correlation between velocity, eccentricity and defect asymmetry. While they are strongest for $v_0 = 1.5$, they decrease for $v_0 = 1.75$ and are almost gone for $v_0 = 2.0$, in accordance with the error bars in Fig. \ref{fig2}(left). 

\begin{figure}[htb]
   \centering
   \includegraphics[width=0.7\textwidth]{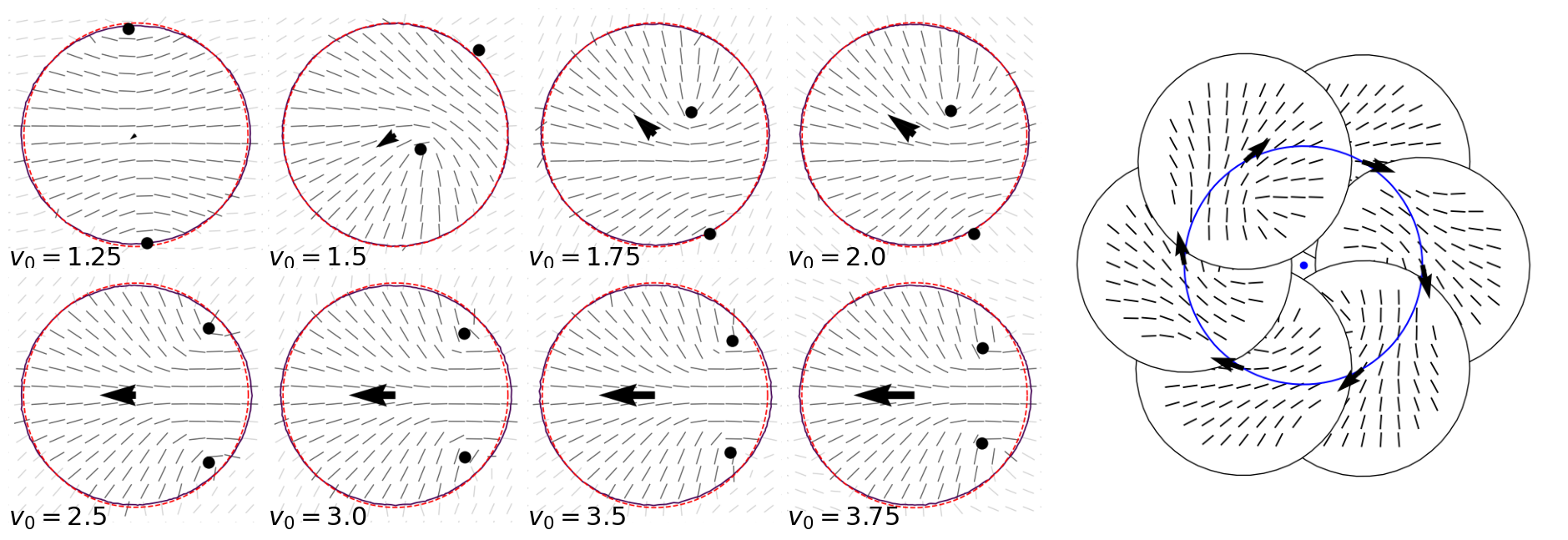}
   \caption{(left) Particle shape, nematic liquid crystal field and position of $+1/2$ defects for various $v_0$ corresponding to the resting regime $v_0 = 1.25$, the spinning regime $v_0 = 1.5, 1.75, 2.0$ and the translation regime $v_0 = 2.5, 3.0, 3.5, 4.0$. To highlight the particle deformation a circular shape of the same area is plotted with the same center of mass. The arrows indicates the particle velocities. (right) Circular particle path for $v_0 = 2.0$.}
    \label{fig3}
\end{figure}

To further highlight the connection between particle velocity, eccentricity and asymmetry of the defect positions Fig. \ref{fig3}(left) shows the particle shape together with the principle eigenvector of the largest eigenvalue of $\mathbf{Q}$ (director field) and the defect positions for various $v_0$. As the defects can also be located at the $0.01$-levelset, the nematic liquid crystal, which is forced to decay to zero in regions with $\psi \leq 0$ is also shown in the vicinity of the particle. The defects deform the director field and the deformed director field is responsible of the symmetry breaking. For the circular or spinning regime the shape deformation is asymmetric with respect to the direction of movement and the defect asymmetry increases with $v_0$. The direction (up or down) depends on the splitting of the $+1$ defect into two $+1/2$ defects and the resulting shape deformation. For the translation regime the shape is symmetric with respect to the direction of motion and also the defect arrangement is almost symmetric. With increasing $v_0$ the defects are located closer to the symmetry axis and the velocity of movement, which only slightly deviates from the symmetry axis, increases. Fig. \ref{fig3}(right) shows a typical circular path together with the corresponding director field and the velocity in the shown time instances. Due to the small radius of the circulation, which is almost independent on the strength of activity $v_0$, we denote this motion as spinning in the following.   

\section{Collective behaviour}
\label{s:collective}

The behaviour in the resting and translation regimes essentially coincides with that of the polar active phase field crystal model \cite{Alaimoetal_NJP_2016}. This also remains true for the emerging collective behaviour in unconfined and confined geometries, see \ref{a1}. We thus only concentrate on the spinning regime in more detail. First, we characterize the behaviour of interacting spinning particles in unconfined and confined geometries for an intermediate packing fraction of $0.57$. To compute the packing fraction we consider the 0.01-levelset of $\psi$ to determine the area of the particles as $A_N = \int I_{\{\psi>0.15\}} \, d \mathbf{r}$. The area of one particle $A = A_N / N \approx 0.9 d^2$ with $d = 4 \pi / \sqrt{3}$ the lattice distance in the phase field crystal model. This essentially motivates to consider the 0.01-levelset. The packing fraction results as $A_N / |\Omega|$ or $A_N / |\Omega_c|$.

\subsection{Synchronisation in unconfined and confined geometries}

\begin{figure}[htb]
    \centering
    \setlength{\unitlength}{\textwidth}
    \includegraphics[width=0.8\textwidth]{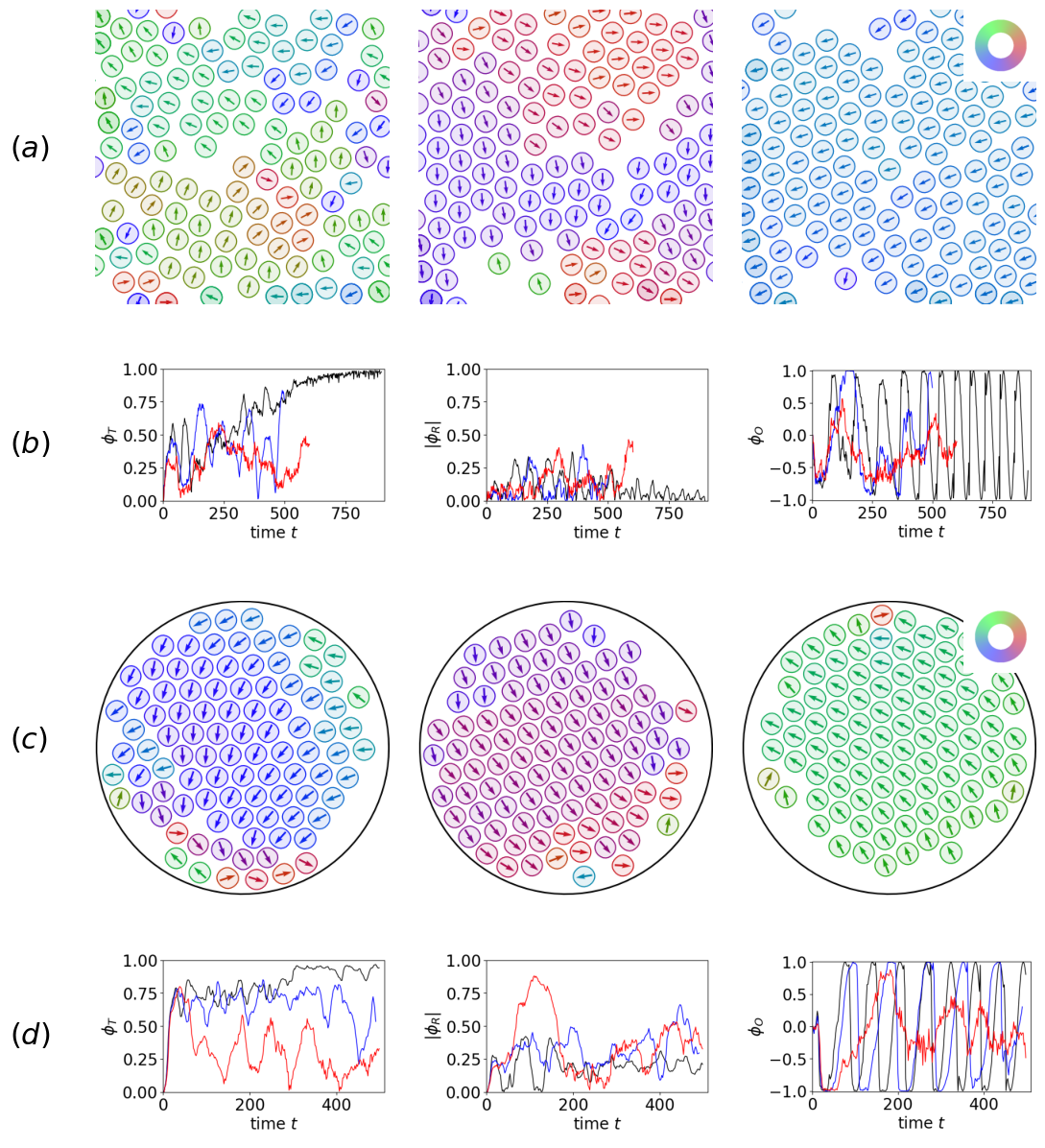}
    \caption{(a) Three different time instances ($t = 100,500,900$ from left to right) indicating the evolution to synchronized spinning for $v_0 = 2.0$ and $N = 120$ in the square domain $\Omega$ with periodic boundary conditions. The particles are visualized by the 0.15-levelset of $\psi$. The color corresponds with the direction of the arrow and indicates the direction of motion. The initial condition is a square lattice of circular particles with perturbed nematic fields. The perturbation leads to a random distribution of the resulting direction of motion. (b) The translational, rotational and angular order parameters (from left to right) for corresponding simulations with $v_0=1.5$ (red), $v_0=1.75$ (blue) and $v_0=2.0$ (black). (c) Three different time instances ($t =  100,300,500$ from left to right) indicating the evolution to synchronized spinning for $v_0 = 2.0$ and $N = 100$ in the circular domain $\Omega_c$. Visualisation and initial conditions are as in (a). (d) The translational, rotational and angular order parameters (from left to right) for corresponding simulations with $v_0=1.5$ (red), $v_0=1.75$ (blue) and $v_0=2.0$ (black). See also Supplementary Movie.}
    \label{fig5}
\end{figure}

We first consider 120 particles in the square domain $\Omega$ with periodic boundary conditions. The self-spinning particles form crystalline structures with local triangular order, with dislocations and regions with no particles, which dynamically rearrange. The particles are self-spinning and due to local interactions some particles also move to positions further away than the spinning radius. This is consistent for all considered self-propulsion strength $v_0$. However, only for $v_0 = 2.0$ the translational order parameter $\phi_T \approx 1$, which indicates translational order or in the current context synchronized spinning. This is confirmed by the angular order parameter $\phi_O$, which oscillates with fixed periodicity, see Fig. \ref{fig5}(a),(b). The behaviour in the circular confinement $\Omega_c$ is similar, see Fig. \ref{fig5}(c),(d). Also in this setting the translational order parameter $\phi_T \approx 1$ for $v_0 = 2.0$ and the angular order parameter $\phi_O$ oscillates with fixed periodicity. This nonequilibrium steady state is reached much faster than in the unconfined geometry. One could conclude that in this setting confinement helps to synchronise the particles. Deviations from synchronized spinning in the reached nonequilibrium steady state are only found at the edge. This corresponds with regions with crystalline defects or no particles. In the centre a crystal with perfect triangular lattice and synchronously spinning particles emerges.    

In contrast to the translational regime considered in \ref{a1} with translational and rotational motion as nonequilibrium steady states, here both settings behave similar. In unconfined and confined geometries the initially independently spinning particles undergo a transition to a nonequilibrium steady state of positional and orientational order, a synchronized spinning crystal. The simulations only confirm this for $v_0 = 2.0$. If this state is also reached at later times for the other self-propulsion strength $v_0$ remains open. 

\subsection{Varying packing fraction and emerging edge currents} 

\begin{figure}[htb]
    \centering
    \includegraphics[width=0.7\textwidth]{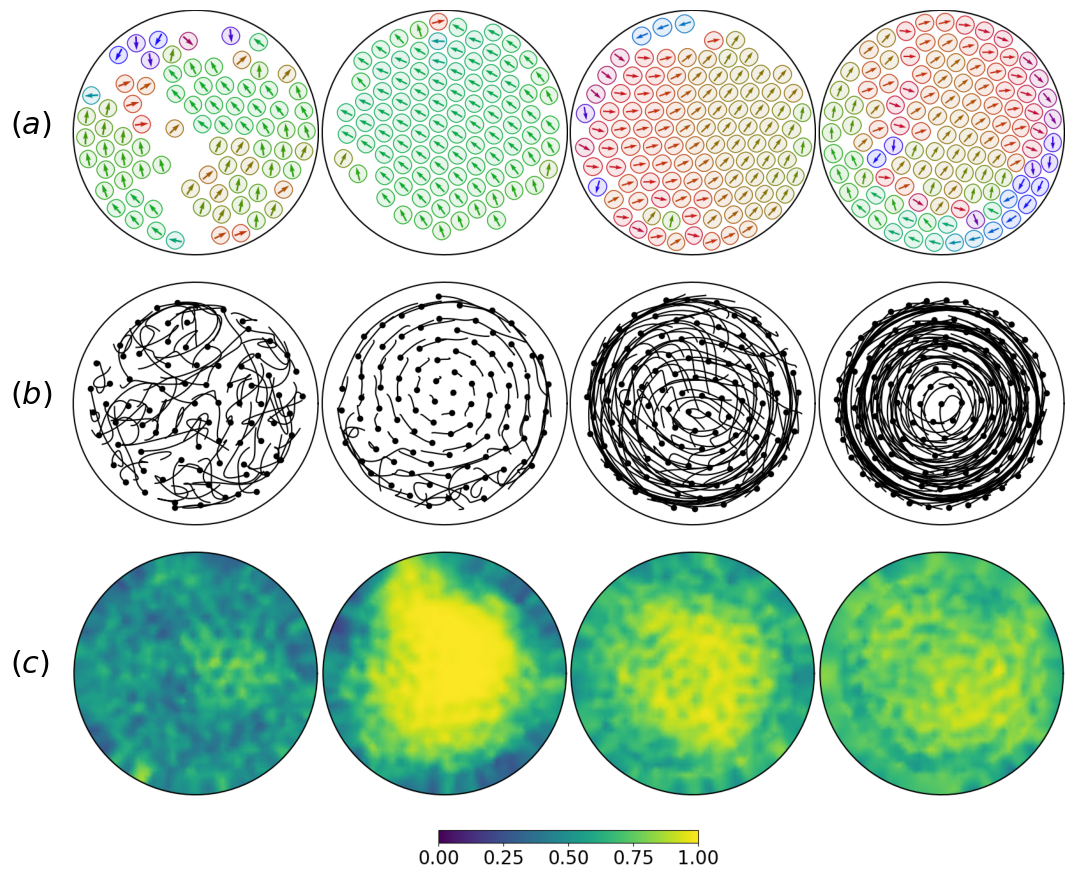}
    \caption{Varying packing fraction $0.48, 0.57, 0.66, 0.7$ (from left to right) with self-propulsion strength $v_0 = 2.0$ in circular domain $\Omega_c$. (a) Time instance $t = 500$. Visualisation and initial conditions as in Fig. \ref{fig5}. (b) Coarse-grained particles trajectories in time interval $(200,500)$. (c) Averaged bond order parameter in time interval $(200,500)$. See also Supplementary Movie.} 
    \label{fig:circleCurved}
\end{figure}

All previous simulations consider a packing fraction of $0.57$. We now vary this in the circular confinement and observe different behaviour, see Fig. \ref{fig:circleCurved}. For a smaller packing fraction of $0.48$, at least within the considered time neither synchronized spinning nor crystal formation can be observed. Instead only small crystalline patches form and dynamically rearrange. Due to the available space particle interactions lead to local positional rearrangements. Similar to the situation in unconfined geometries particles can move to positions further away than the spinning radius. Fig. \ref{fig:circleCurved}(b) shows the coarse-grained trajectories of the particles (without the spinning component), and Fig. \ref{fig:circleCurved}(c) the bond number averaged over a larger time frame. The chaotic trajectories and the low bond number for a packing fraction of $0.48$ underpin the described behaviour. The bond number gives an indication of crystalline order and is computed for particle $j$ as $b^n_{6j} = (\sum_{k \in N_j} e^{6i \theta^n_{jk}}) / N_j$, with $N_j$ the nearest neighbors of particle $j$ within a specified radius related to $d$ and $\theta^n_{jk}$ the angle between $\mathbf{r}^n_k - \mathbf{r}^n_j$ and the x-axis. The considered averaged bond number $\hat{b}_{6j}$ accounts for the average over various times $t^n$. $\hat{b}_{6j} = 0$ considers the situation of an isolated particle and $\hat{b}_{6j} = 1$ that of a perfect triangular lattice, a particle with six neighbors. The nearest neighbors are constructed using a Voronoi-tesselation for the centers of mass. For the low packing fraction the system is in a fluid like regime with isolated particles which can easily change their positions. 

For packing fraction $0.57$ the coarse-grained trajectories show more or less stationary particles in the center and only small movements on the edge, see Fig. \ref{fig:circleCurved}(b). This small edge currents differ from the behaviour in the unconfined geometry discussed above. The emerging edge currents have an effect on the crystalline structure, which is quantified by the averaged bond number, see Fig. \ref{fig:circleCurved}(c). With $\hat{b}_{6j} \approx 1$ it shows a clear persistent triangular lattice in the center and deviations only at the edge. 

The situation changes for increasing packing fraction. For $0.66$ and $0.7$ the dominating situation of a crystal with triangular lattice and synchronously spinning particles is destroyed. The edge currents increase and propagate towards the center, see Fig. \ref{fig:circleCurved}(b). While for $0.66$ a triangular lattice still exist at least over some time span before it gets rearranged, the averaged bond order for $0.7$ has even less indication of such stable crystalline order, see Fig. \ref{fig:circleCurved}(c). The coarse grained particle trajectories, see Fig. \ref{fig:circleCurved}(b), show a transition towards a global vortex. The snapshots in Fig. \ref{fig:circleCurved}(a) further indicate that the particles no longer spin synchronously. The Supplementary Movies further confirm this behaviour.

\begin{figure}[htb]
    \centering
    \includegraphics[width=0.9\textwidth]{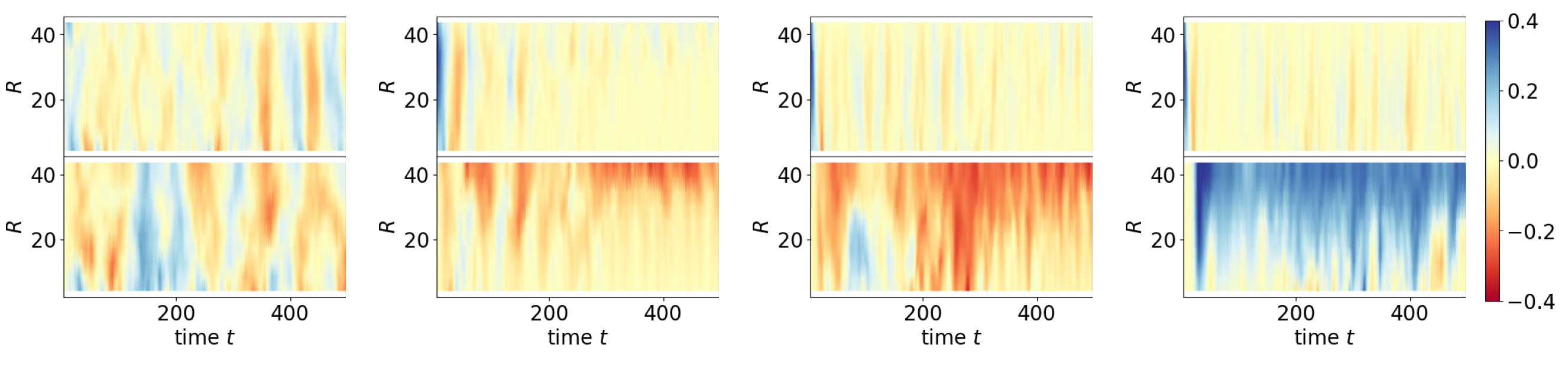}
    \caption{Kymographs corresponding to simulations in Fig. \ref{fig:circleCurved}. (top) Orthoradial component and (bottom) radial component of the particle velocity averaged over all particles with distance $R$ from center for varying packing fraction  $0.48, 0.57, 0.66, 0.7$ (from left to right) with self-propulsion strength $v_0 = 2.0$.} 
    \label{fig:kymograph}
\end{figure}

The coarse-grained particle movements (without the spinning component) is further analysed in Fig. \ref{fig:kymograph}, which confirms the above discussion. The kymographs show the orthoradial and radial components of the coarse-grained velocity averaged over all particles which are located at a distance $R$ from the center of the domain $\Omega_c$. While there is almost no movement in orthoradial direction, the slight edge currents for packing fraction $0.57$ in the radial component and their increased strength and extension towards the center for packing fractions $0.66$ and $0.7$ are clearly visible. The direction of the emerging vertex rotation depends on the shape of the particles at the boundary as a result of their spontaneous symmetry breaking. The majority decides on the emerging direction at the edge, which persists towards the center.

\section{Discussion}
\label{s:collective}

The proposed minimal model allows to explore different dynamical regimes by varying one activity parameter only. The direct coupling of the self-propulsion strength $v_0$ with the internal nematic structure and the deformability of the particle leads to slightly deformed resting states if $v_0$ is below some threshold. It induces within a certain parameter range chirality, which leads to circular or spinning motion. If it is above some threshold a symmetric shape and translational motion emerges.  
All regimes have been investigated in unconfined and confined geometries. While the translational regime is more or less identical with the behavior of the active polar phase field crystal model \cite{Alaimoetal_NJP_2016} and the observed rotational behavior in circular confinements reminiscent of various experiments, e.g., on highly concentrated bacterial suspension which self-organize into a single stable vortex \cite{Wiolandetal_PRL_2013}, collective behavior of self-circulating or self-spinning particles are much less explored. Self-spinning particles are computationally considered in circular confinements \cite{vanZuidenetal_PNAS_2016}. While fundamental issues differ, e.g., our particles are deformable, our spinning radius is significantly larger and our spinning velocity significantly lower, the emerging behaviour is similar. The competition between circular confinement, self-propulsion and steric interactions can lead to the emergence of edge flows and rotations. Within a wider perspective, similar edge flows and rotations have also been observed in chiral fluids \cite{Sonietal_NP_2019}. It is shown that in systems of synchronously spinning particles both parity (or mirror) symmetry and time-reversal symmetry are broken. Hydrodynamic theories with additional terms to account for these broken symmetries, e.g., rotational viscosity tend to force the fluid as a whole to rotate with the angular velocity of the spinning particles. However, the motion of the fluid in the bulk is suppressed by friction. As a result, the fluid moves mostly at the boundary and the penetration depth of the vorticity of the fluid from the boundary into the bulk is controlled by the shear viscosity. These results are similar to the edge currents in our simulations and their propagation towards the center with increasing packing fraction. These similarities with other systems which range from collective rotation of chiral molecules of a liquid crystal \cite{Nitonetal_Nano_2013}, to sperm cells \cite{Riedeletal_Science_2005,Friedrichetal_PNAS_2007}, colloidal and millimeter scale magnetes \cite{Grzybowskietal_Nature_2000,Grzybowskietal_PNAS_2002} and rotating robots \cite{Scholzetal_NC_2018}. An interesting biological example is provided by Chlamydomonas reinhardtii (C. reinhardtii). This micron-sized unicellular algae is able to self-propel to perform translational motion, but also has the ability to self-rotate \cite{Ravazzanoetal_SM_2020}. Rotation is used to sense the direction of light to optimize efficiency of phototaxis \cite{Choudharyetal_BPJ_2019}.

\begin{figure}[htb]
   \centering
   \includegraphics[width=0.8\textwidth]{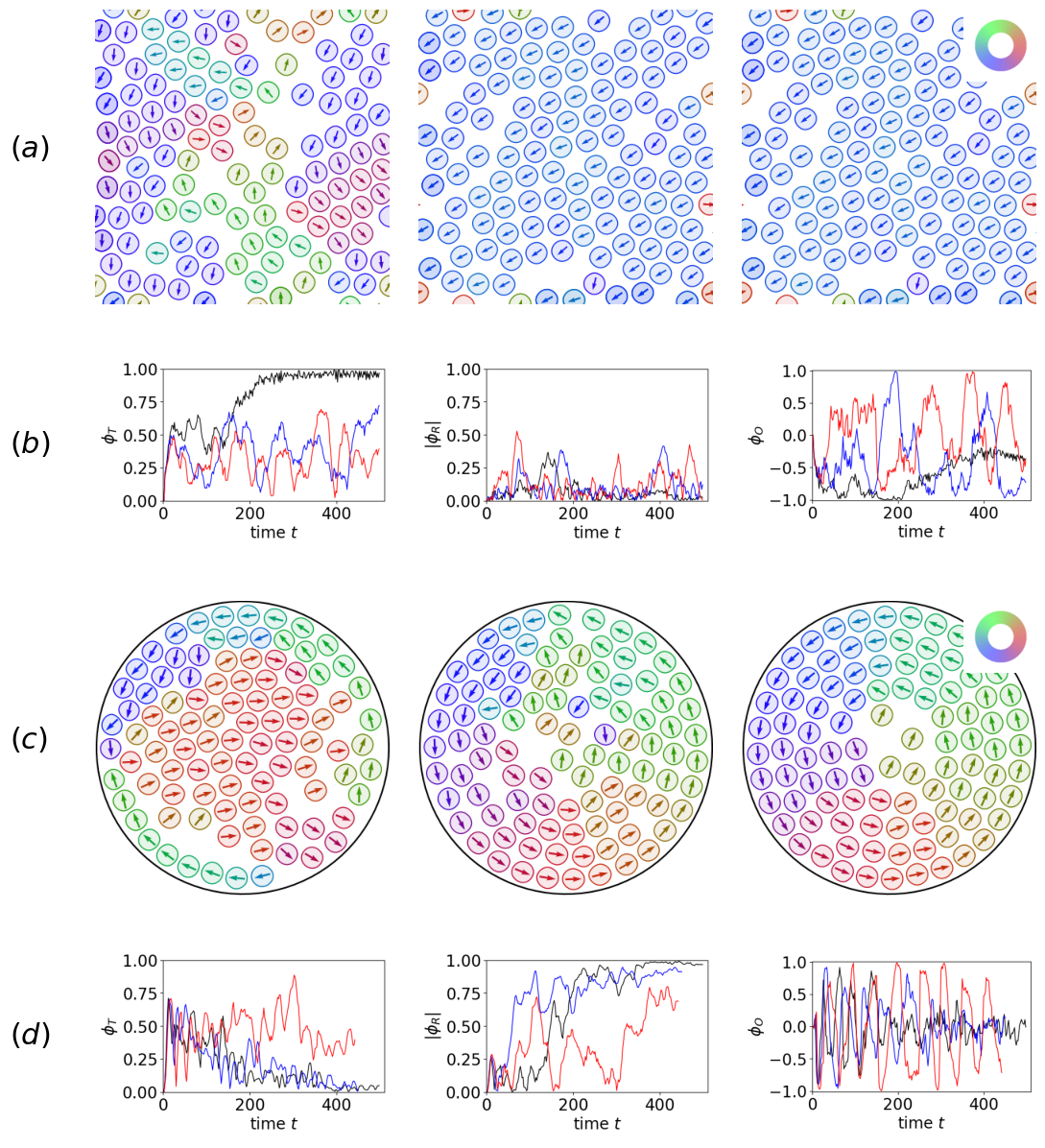}
   \caption{(a) Three different time instances ($t = 100,300,500$ from left to right) indicating the evolution to collective migration for $v_0 = 4.0$ and $N = 120$ in a square domain $\Omega$ with periodic boundary conditions. Visualisation and initial conditions are as Fig. \ref{fig5}. (b) The translational, rotational and angular order parameters (from left to right) for corresponding simulations with $v_0=3.0$ (red), $v_0=3.5$ (blue) and $v_0=4.0$ (black). (c) Three different time instances ($t = 100,300,500$ from left to right) indicating the evolution to collective migration for $v_0 = 4.0$ and $N = 100$ in the circular domain $\Omega_c$. Visualisation and initial conditions are as in (a). (d) The translational, rotational and angular order parameters (from left to right) for corresponding simulations with $v_0=3.0$ (red), $v_0=3.5$ (blue) and $v_0=4.0$ (black). See also Supplementary Movie.}
    \label{figa1}
\end{figure}

The proposed microscopic field theoretical model can be extended towards various directions, e.g., hydrodynamic interactions. This is already considered within the phase field crystal model for passive systems, e.g. \cite{Goddardetal_JPCM_2013,Tothetal_JPCM_2014,Praetoriusetal_JCP_2015,Heinonenetal_PRL_2016}.
Other possibilities consider multicomponent systems \cite{Elderetal_PRB_2007,Alaimoetal_PRE_2018}.

\appendix

\section{Collective behaviour in translational/rotational regime}
\label{a1}

If the self-propulsion strength is above some threshold, a symmetric arrangement of defects and a symmetric shape of the particles is enforced. As a result, the particles behave as in the polar active phase field crystal model \cite{Alaimoetal_NJP_2016}. This behaviour leads to collective translation in unconfined geometries and collective rotation in circular confinements, see Fig. \ref{figa1}. It shows simulations for different values of $v_0$. Collective behaviour is only reached in the considered simulation time for the largest values, $\varphi_T \approx 1$ for $v_0 = 4.0$ and $|\varphi_R| \approx 1$ for $v_0 = 3,5$ and $4.0$. This behaviour is in qualitative agreement with results in \cite{Alaimoetal_NJP_2016} and corresponding large scale simulations of multiphase-field models \cite{Praetoriusetal_NIC_2017}.  

\ack
This work is funded by German Research Foundation within project FOR3013. We used computing resources provided by JSC within project HDR06.

\section*{Bibliography}
\bibliographystyle{unsrt} 
\bibliography{lib}

\end{document}